\documentclass[10pt]{article}

\usepackage[utf8]{inputenc}
\usepackage[T1]{fontenc}
\usepackage{amsmath}
\usepackage{amssymb}
\usepackage{graphicx}
\usepackage{natbib}
\usepackage{xcolor}
\usepackage{hyperref}
\begin{document}

\title{Modelling Palomar Transients: Constraints from Reflection Geometry and Orbital Altitude}

\author{
Beatriz Villarroel\thanks{Nordita, KTH Royal Institute of Technology and Stockholm University, Hannes Alfv\'ens v\"ag 12, SE-106 91 Stockholm, Sweden; e-mail: beatriz.villarroel@su.se}
\and
Alina Streblyanska\thanks{Society of UAP Studies}
\and
Hichem Guergouri\thanks{Science of the Matter Division, Research Unit in Scientific Mediation, CERIST, Constantine, Algeria}
\and
Brian Doherty\thanks{Independent researcher}
\and
Matthew Shultz\thanks{Independent researcher}
\and
Stephen Bruehl\thanks{Department of Anesthesiology, Vanderbilt University Medical Center, 701 Medical Arts Building, 1211 Twenty-First Avenue South, Nashville, TN 37212, USA}
}

\date{}

\maketitle

\begin{abstract}
Recent searches of digitised photographic plates from the Palomar Observatory have uncovered tens of thousands of short-lived transients, each visible in only a single exposure. Their morphologies indicate sub-second flashes, consistent with specular reflections from highly reflective objects in near-Earth orbits.

In this paper, we present a modelling study using geometric shadow modelling, Monte Carlo simulations, and photometric constraints to infer their physical properties.

We examine the observed groupings and alignments, which appear to be consistent with sudden changes in attitude.

Assuming a spherical-shell model, the angular profile of the measured transient deficit around the antisolar point is consistent with a population at characteristic altitudes of $\sim 20{,}000$--$25{,}000$ km. A complementary estimate, based on the altitude dependence of the global Earth-shadow deficit, yields a broader characteristic range of $\sim 20{,}000$--$35{,}000$ km above Earth's surface, extending into the geosynchronous orbital (GSO) region.

Under simplified geometric and photometric assumptions and assuming $\sim20{,}000$ to $35{,}786$ km above the Earth's surface, the inferred sizes of the specularly reflecting facets range from centimetre scales up to $\sim3$ m, with characteristic flash durations of $\sim320$ ms and slow rotation rates. We explore both natural and non-natural toy models consistent with the data, and provide constraints to guide future observational searches.
\end{abstract}


\section{Introduction}
When we look up at the night sky, it is far from static. It is populated by a rich variety of transient light sources: comets and asteroids drift through our field of view, variable stars change their brightness on timescales ranging from hours to years, and flaring stars and active galactic nuclei can produce sudden, short-lived bursts of light. In the last seventy years, however, the sky has acquired an additional layer of variability. It is now increasingly filled with artificial transients -- brief glints and streaks produced by satellites and space debris -- many of which are visible to the naked eye. Their numbers continue to grow rapidly year by year.

For astronomers interested in searching for brief glints and transients associated with either non-anthropogenic artificial objects or yet undiscovered natural objects, old photographic plates taken before the launch of the first artificial satellite constitute invaluable archives, preserving a view of the night sky untouched by humans. These plates, typically large glass plates coated with light-sensitive photographic emulsions, record astronomical sources across different wavelengths, depending on the emulsions and filters employed, and have been used at observatories all over the planet to survey the night sky. A remarkable body of foundational astronomy has been built upon these plates, including the discovery of the expansion of the Universe. Thanks to the digitisation of historical photographic plates, a population of unusual transient objects has in recent years been identified in images from Palomar \citep{Villarroel2020,Villarroel2021}. These sources appear and vanish within a single plate exposure, implying extremely short lifetimes on the order of seconds or less. 
Such events were largely dismissed and overlooked by earlier generations of astronomers, on the grounds that they were plate defects, particularly as they do not recur in subsequent images. There is no doubt that a significant fraction of point sources detected on historical photographic plates are attributable to defects, particularly those clustering toward the plate edges. Artefacts such as bubbles, dust, or small imperfections in the emulsion can, under certain conditions, are claimed to mimic point-like stellar sources \citep{Greiner1990,Hambly2024}.



The Palomar transients were first reported by \cite{Villarroel2020}. They exhibit statistical properties that are difficult to reconcile with local plate defects. Their light profiles are star-like, albeit slightly narrower than those of typical stars of comparable brightness. These more compact and circular point-spread functions (PSFs) were noted by \cite{Hambly2024}, who interpreted them as emulsion defects. However, such brightness profiles are also fully consistent with brief flashes lasting from sub-second to maximum of a few seconds \citep{VillarroelARXIV,Busko2026}. 

More intriguingly, the transients sometimes appear and vanish in groups. Two notable examples are a group of nine transients distributed across a $10 \times 10$ arcmin$^2$ region \citep{Villarroel2021}, and a triple transient reported by \cite{Solano2024}. If the grouped events are physically associated and synchronised within $\sim 30$ min, the finite speed of light constrains their distance to $\lesssim 0.02$ light-years for an angular extent of 10 arcmin, placing the phenomenon on Solar-System scales (see Section A.8 in the Supplementary Information of \citealt{Villarroel2021}). This unexpected finding led some scientists to propose that the transients might instead be scanning artifacts or plate defects that could be identified through close inspection, for example by microscopy \citep{Hambly2024}. A close inspection of the nine transients on the original glass plates, together with the triple-transient case on a copy plate, confirms the morphologies inferred from the digitised data \textcolor{blue}{(Yang \& Villarroel, submitted)}.

Similar short flashes are today known to result from specular reflections off satellites and space debris in geosynchronous orbit \citep{Nir2020}. In some cases, transients appear to form narrow alignments or band-like structures \citep{VillarroelPASP}, with probabilities of alignment that are statistically unlikely under random expectations but consistent with predictions for artificial objects producing specular solar reflections \citep{Villarroel2022a}.

However, it is at this point that some particularly intriguing and unexpected statistical patterns begin to emerge. The transients exhibit temporal correlations at the 3$\sigma$ level with nuclear weapons tests and reported UFO events \citep{Bruehl2025}. Even more strikingly, the transients the transients appear to diminish significantly in number in the Earth's shadow \citep{VillarroelPASP}. A shadow deficit of $\sim$30\% is observed, corresponding to a significance of $7.6\sigma$ when compared to the expected fraction of sky lying within the shadow during the observations.

In a new analysis led by \cite{Bruehl2026}, we show that both the transient–nuclear test correlation and the Earth-shadow deficit strengthen systematically as each transient is assigned a probability of being a real phenomenon (rather than an instrumental artifact) using a machine-learning model trained on visually classified data. As progressively more ambiguous contaminants are removed, both effects increase in magnitude and statistical significance. Such behaviour would not be expected if the signals were dominated by plate defects, but instead provides strong evidence that they reflect an authentic physical phenomenon.

The main analytical results regarding transient alignments, the shadow deficit, and transient-nuclear testing association have been independently replicated, and in some cases extended, by independent researchers. The raw-detection stage of the VASCO transient-search pipeline has been independently reproduced, recovering approximately $94$--$98\%$ of the catalogue sources \citep{Ahlberg2026}\footnote{J. Ahlberg emphasises the importance of using full-plate scans for reproducibility. Further practical recommendations for independent replication efforts are provided in a blog post by the VASCO Network \citep{VASCOBlog2026}.} The identification of transient groupings in the Palomar data has also been confirmed by \citet{Hayes2026}, using an independent transient-detection pipeline. Using the VASCO main transient sample of 107,875 transients, the main statistical findings related to the transient–nuclear test correlations have likewise been reproduced by \cite{Doherty2026}, \cite{Cann2026a}, and \textcolor{blue}{Sinkkonen (2026)}\footnote{https://github.com/euxoa/plates}. The Earth-shadow deficit has also been independently confirmed by \cite{Doherty2026}, who further found that the transient–nuclear correlation becomes stronger when restricting the sample to transients in sunlit regions (and thus potentially more likely to represent reflective objects).

Further unexpected patterns have emerged from independent analyses. \cite{Cann2026a,Cann2026b} reported a pronounced anticorrelation between transient detection rates and the geomagnetic storm index. This is totally inconsistent with transients being the result of solar radiation effects in the Earth's atmosphere. It also cannot be explained as a simple consequence of reduced detection efficiency during auroral activity due to increased sky brightness \citep{Cann2026c}. \cite{Doherty2026b} also found that the high-confidence transients (as identified using machine learning) avoid the ecliptic plane, making asteroids, comets, and zodiacal debris unlikely as the dominant source population. The latter study further identified a strong excess of close pairs (duplets) and rare triplets among the transients, as well as a significant correlation between close-pair excess and linear alignments. Intriguingly, a triple transient candidate has previously been reported by the VASCO team on a POSS-I plate exposed on 19 July 1952, temporally coincident with the Washington 1952 UFO flap \citep{Solano2024,VillarroelPASP}.

But most striking is that similar transients exhibiting comparable morphologies and slightly sharper profiles consistent with specular reflections, have also been identified in the Hamburg plate archives taken from the same pre-Sputnik time period \citep{Busko2026}. This rules out instrumental defects or scanning errors specific to the Palomar plates. Furthermore, \citet{Busko2026b} showed that transient sources on historical photographic plates exhibit the same optical coma distortions as nearby stars, demonstrating that the signals originated from light passing through the telescope optics rather than from plate defects. The study also reported a coma-distorted triple transient, providing independent support for the triple event described by \citet{Solano2024}. Overall, the evidence points towards these objects representing genuine detections of sources located beyond the atmosphere.

Observations of transient sources beyond the atmosphere become particularly interesting in light of newly released \textit{Redstone-Mercury} and \textit{Apollo} mission documents, which contain repeated astronaut reports of flashing objects observed outside the spacecraft windows and not easily attributable to prosaic explanations. Several accounts describe slowly tumbling or rhythmically flashing objects, in some cases appearing as flat or angular fragments with apparent structure. Other reports describe more distant objects exhibiting regular flashing patterns, apparent rotation, and even alternating double flashes consisting of a bright flash followed by a dimmer one, distinct from nearby particles or debris. The sizes and distances remained uncertain, although some astronauts estimated dimensions on the order of several inches \citep{Apollo11Debrief1969,Apollo12Transcript1969,Apollo17Transcript1972,SkylabDebrief1973}. The documents are publicly available through the U.S. government archive \footnote{https://www.war.gov/ufo}. Interestingly, one of the released \textit{Apollo}-era images appears to show three luminous objects above the lunar horizon \citep{Apollo17Transcript1972}.

This papers proceeds from the assumption that the transients originate from real sources in Earth orbit, and aims to apply simple toy models to their statistical properties and thereby infer the range of altitudes, geometries, motions, and linear dimensions of the underlying reflective objects. We restrict ourselves to explanations not yet ruled out by previous papers \citep{Villarroel2021,VillarroelPASP,Bruehl2025,Bruehl2026}.

This analytical approach using various toy models, preliminary and subject to further refinement as the transients are studied in greater detail, will be used to make predictions and inform the design of an observational campaign aimed at detecting such transients in the present day. 

In Section \ref{sec:Methods}, we describe the transient sample used in the analysis. In Sections \ref{sec:Results} and \ref{Constraints}, we present the modelling and the inferred properties of the objects. In Section \ref{sec:Future}, we outline possible next steps and discuss inferred properties of the transients that have been observed, including where and what to search for, providing practical recommendations for modern-day observations.  Finally, in Section \ref{sec:Conclusions}, we summarize the main findings of the paper.

\section{Methods}\label{sec:Methods}

\subsection{Samples}
\subsubsection{Main sample}
The Palomar Observatory Sky Survey I (POSS-I) was one of the first comprehensive photographic surveys of the northern sky, conducted between 1949 and 1958 using the 48-inch Samuel Oschin Schmidt telescope at Palomar Observatory. The survey used large-format photographic plates (approximately $6^\circ \times 6^\circ$) with emulsions sensitive primarily to blue (O-band) and red (E-band) wavelengths, reaching limiting magnitudes of roughly $R \sim 20$ in red. Typical exposure times were on the order of 40--60 minutes.

We use a sample of 107,875 transients from POSS-I, based on the filtering process described in \cite{Solano2022}. Clarifications on how the original sample of more than 298,000 transients was filtered down to 107,000+ sources are described in \cite{villarroelCommentary}\footnote{Requests for the sample or additional questions should be addressed to Enrique Solano.}. This sample contains a substantial fraction of false positives e.g. plate defects and stars; see Section 2 of \cite{VillarroelPASP}. Approximately one-third of the sample is estimated to arise from solar reflections, as inferred from the deficit of transients within the Earth's shadow. Since emulsion defects do not follow the time-dependent Earth–Sun geometry and stars lie far beyond the Earth's shadow, the background contamination may be considered effectively constant, with no impact on the shadow geometry.

The main sample is further analysed using machine learning in \cite{Bruehl2026}, where each transient is assigned a probability score ($p$) to distinguish genuine transient candidates from scratches, plate defects, and ordinary stars. Whenever the available statistics permit, we use this more refined sample.

\subsubsection{Control sample}
We construct a control sample using the central coordinates and observation dates and times from the STScI list of XE plates. Only those red plates (XE-series) included in this study, i.e.\ plates on which any of the 107,875 transients were detected, are considered. This corresponds to a total of 635 plates. For each plate, we generate 100 random points within the field, assuming a plate size of approximately $6 \times 6$ degrees centred on the reported coordinates. The control sample allows us to compare the results while accounting for the actual observing times and sky coverage.

\subsection{The antisolar distance}\label{sec:earthshadow}

We determine whether a transient falls within Earth’s shadow using the dedicated software library \textit{earthshadow} developed by \cite{Nir2024_git}.  This code has been extensively applied in our previous works and is described in detail in \cite{VillarroelPASP}. In brief, the code evaluates whether a point at a given altitude and geographic position lies within Earth’s shadow by modeling the Sun–Earth–object geometry and the solar illumination angle. It accounts for the altitude-dependent angular radius of Earth’s shadow. For the purpose of this initial analysis, we first assume an altitude corresponding to geosynchronous orbit (GSO), i.e., $r \approx 42{,}164$ km including Earth's radius (or $35{,}786$ km above Earth's surface). This assumption is subsequently changed.

\begin{figure*}
\includegraphics[width=\textwidth]{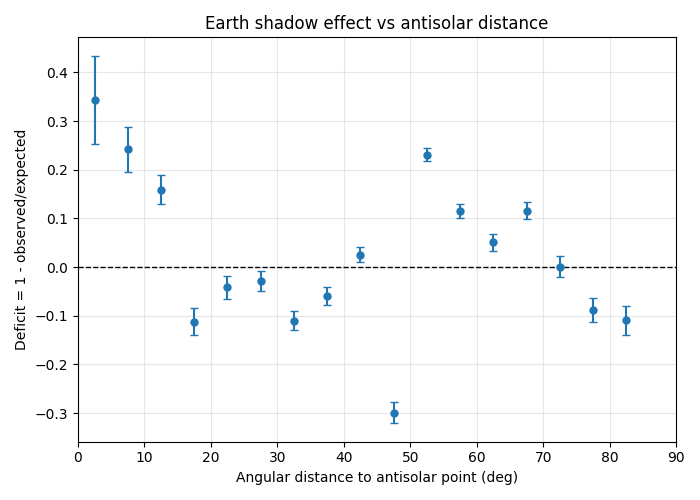}
\caption{\label{deficit1} {\bf Deficit of transients in the shadow.} We show the deficit of transients among the 107,875 transients as a function of angular distance from the antisolar point. Larger positive values indicate a stronger deficit relative to the control sample. A shadow-related signal is expected to be strongest near the antisolar point ($\theta \approx 0^\circ$) and to decrease towards zero at larger angular distances. The observed profile initially follows this expectation, while becoming more complex at larger angular distances, where additional effects may contribute.}
   \end{figure*}

We run the code for our transient sample and control sample, and compare the observed fraction of transients in shadow against expectations from the control sample. The results clearly show a deficit of transients in the Earth’s shadow as a function of angular distance from the antisolar point. An example for geosynchronous altitude is shown in Figure \ref{deficit1}. We also attempt this exercise across nearby altitudes, ranging from 20{,}000 km -- 80{,}000 km.

The angular profile indicates a deficit that decreases with increasing angular distance, consistent with geometric suppression inside the Earth’s shadow. Specifically, the largest shadow deficit is observed in regions closest to the antisolar point. At larger angular distances, additional effects likely begin to influence the observed profile, suggesting that the signal does not arise from a single physical mechanism alone. A full physical forward model would require knowledge of the orbital distribution, object geometries, rotational states, reflectivities, phase-angle effects, and possible refraction-related scattering properties of the transient population, all of which remain unknown at present. Nevertheless, simple shadow geometry alone predicts structure in the vicinity of the antisolar point at angular scales comparable to those observed. In this work, we therefore focus primarily on the shadow-related deficit near the antisolar point, while more detailed modelling of additional illumination and scattering effects is deferred to future study. A future study will also challenge the spherical-shell model and explore alternative spatial distributions, such as a ring-like structure near the equatorial plane.

To further investigate the physical and orbital properties of the transient sources, we apply forward-modelling and Monte Carlo methods combining geometric shadow modelling, photometric constraints, and rotational simulations.

\subsection{Altitude estimation from the global deficit}\label{sec:globaldeficit}

A second, independent way to determine whether a deficit exists is to investigate its size as a function of orbital altitude. The characteristic orbital altitude will produce the largest and most significant deficit in the Earth’s shadow.

We define the deficit as
\begin{equation}
D = 1 - \frac{f_{\mathrm{obs}}}{f_{\mathrm{control}}},
\end{equation}
where $f_{\mathrm{obs}}$ and $f_{\mathrm{control}}$ are the fractions of objects falling within the corresponding Earth-shadow region.

We recompute the global Earth-shadow deficit for different assumed altitudes using the full transient catalogue and the control sample. The Earth-shadow deficit is highly statistically significant at all tested altitudes, with the largest and formally most significant deficit occurring at an assumed altitude of 25,000 km.

\begin{table}[ht]
\centering
\caption{textbf{Earth-shadow deficit for different assumed orbital altitudes using the complete and quality-filtered (Q-filtered) transient samples.} The preferred characteristic altitude is shown in bold.}
\label{tab:global_shadow_deficit}

\resizebox{\textwidth}{!}{%
\begin{tabular}{lcccccc}
\hline
Sample
& Altitude above surface
& Geocentric shell radius
& Observed shadow
& Control shadow
& Deficit
& Poisson error \\
& (km) & (km) & & & & \\
\hline
Complete
& 20,000      & 26,371 & 1085 & 824 & 0.435 & 0.026 \\
& \textbf{25,000} & \textbf{31,371} &\textbf{ 372}  & \textbf{374} & \textbf{0.573} & \textbf{0.031} \\
& 35,000      & 41,371 & 185  & 182 & 0.564 & 0.046 \\
& GSO: 35,786 & 42,164 & 172  & 153 & 0.517 & 0.054 \\
& 50,000      & 56,371 & 156  & 127 & 0.473 & 0.063 \\
& 55,000      & 61,371 & 146  & 115 & 0.455 & 0.068 \\
& 60,000      & 66,371 & 130  & 103 & 0.458 & 0.071 \\
\hline
Q-filtered
& 20,000      & 26,371 & 196 & 824 & 0.504 & 0.039 \\
& 25,000      & 31,371 & 44  & 374 & 0.755 & 0.039 \\
& \textbf{35,000} & \textbf{41,371} & \textbf{17}  & \textbf{182} & \textbf{0.805} & \textbf{0.049} \\
& GSO: 35,786 & 42,164 & 17  & 153 & 0.768 & 0.059 \\
& 50,000      & 56,371 & 17  & 127 & 0.721 & 0.072 \\
& 55,000      & 61,371 & 17  & 115 & 0.692 & 0.080 \\
& 60,000      & 66,371 & 17  & 103 & 0.656 & 0.090 \\
\hline
\end{tabular}%
}
\end{table}

We repeat the same analysis using the machine learning-cleaned transient sample from \cite{Bruehl2026}.  To minimise systematics related to plate defects, we use the machine-learning filtered sample described in \cite{Bruehl2026} and select only high-probability transients ($p > 0.7$). We note that the empty stripes present in the spatial distribution may influence the results, as may edge effects near the boundaries of the transient surveys. We refer the reader to \cite{Solano2022} to learn how plate edges were treated. We therefore exclude the lowest declinations corresponding approximately to one plate radius, restricting the analysis to $\mathrm{Dec.} > 3^{\circ}$, and focus on the central region defined by $100^{\circ} < \mathrm{R.A.} < 250^{\circ}$. The resulting samples (``quality-filtered samples'') contain 11,418 transient candidates and 23,814 control objects.

The Earth-shadow deficit becomes stronger at all tested altitudes after applying the quality filtering. The largest deficit is found at an assumed altitude of 35,000~km, where it reaches $0.805 \pm 0.049$ ($16.4\sigma$), and is therefore itself highly statistically significant. The formally most significant deficit occurs at 25,000~km ($19.4\sigma$), partly because the larger number of objects within the corresponding shadow region provides greater statistical power than at 35,000~km. Statistical significance therefore cannot by itself be used to identify the characteristic altitude, as it depends on sample size. For this comparison, the absolute magnitude of the deficit is the more relevant interpretive criterion. 

At GSO altitude, the deficit increases from $0.517 \pm 0.054$ in the full sample to $0.768 \pm 0.059$ in the quality-filtered sample. The systematic strengthening of the Earth-shadow deficit as the sample is cleaned indicates that the rejected candidates dilute, rather than produce, the observed signal, consistent with \citet{Bruehl2026}.

For the quality-filtered sample, we also reconstructed the angular deficit profile using different numbers of bins (15, 20, 25, and 30) to examine whether the observed structure is sensitive to the choice of binning. Its profile remains stable, showing a large deficit close to the antisolar point, followed by a transition region and a recovery at larger angular distances. This indicates that the feature is not produced by a particular choice of bin size.

Taken together, the exercise favours a characteristic orbital altitude of approximately $25,000$--$35,000$~km above the Earth's surface, extending into the GSO region. The results are shown in Table \ref{tab:global_shadow_deficit}.

\subsection{Declination distribution of high-probability transients}

We further examined the distribution of the high-probability transient sample as a function of declination to determine whether the candidates exhibit a preferred spatial distribution relative to the celestial equator. This is of interest, as modern optical glints are predominantly reported from objects near the equatorial plane, corresponding to low declinations (see e.g. \citealt{Nir2020}).

\begin{figure*}
    \centering
    \includegraphics[width=0.82\linewidth]{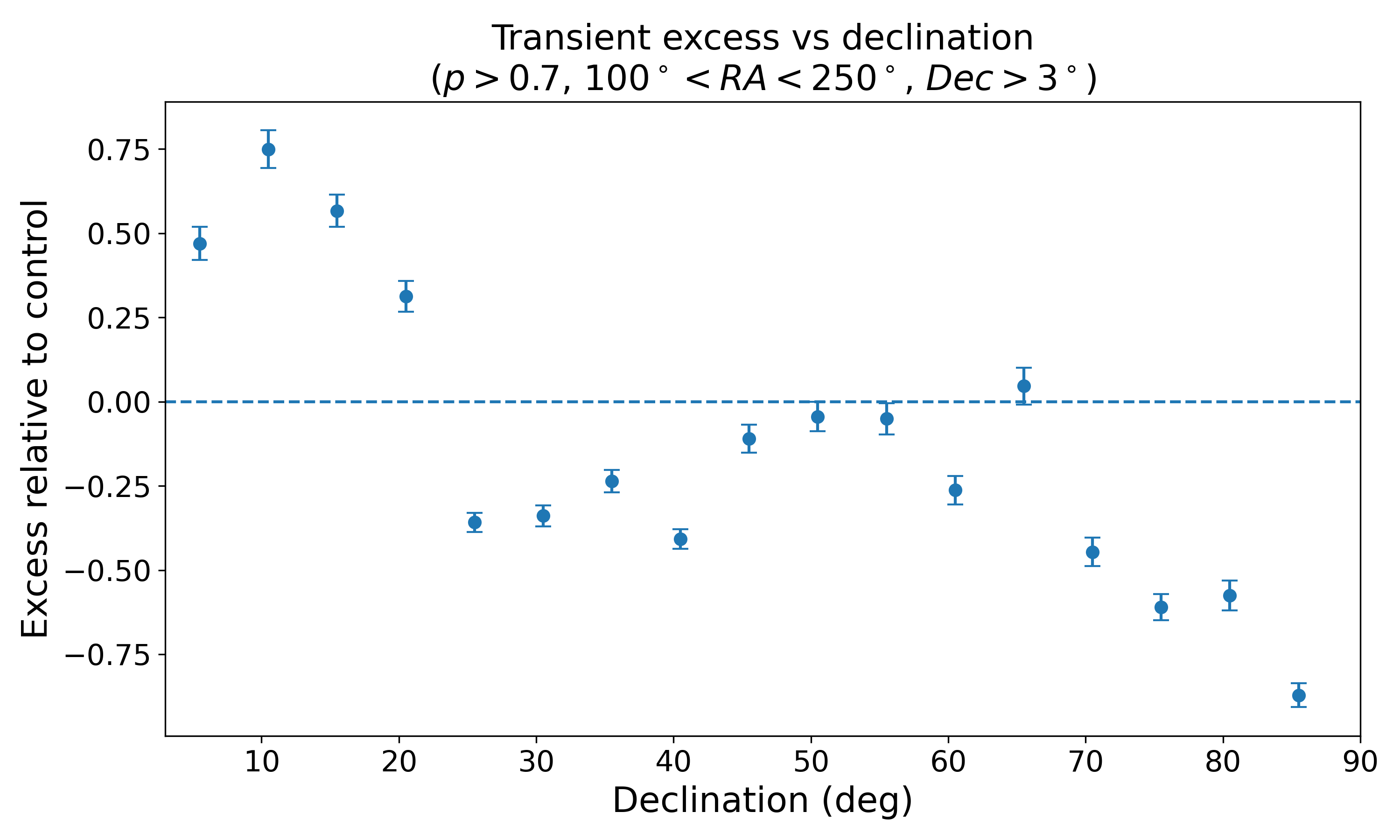}
    \vspace{-2mm}
    \caption{\textbf{Relative excess of high-probability transients as a function of declination, compared to the control sample.} The analysis is restricted to transients satisfying $p > 0.7$, $100^{\circ} < \mathrm{R.A.} < 250^{\circ}$, and $\mathrm{Dec.} > 3^{\circ}$. A pronounced excess is observed at low declinations, approximately corresponding to regions near the equatorial plane. A second excess is present around $\mathrm{Dec.} \sim 65^{\circ}$, followed by a decline towards higher declinations. The origin of this secondary peak is currently unclear. The total number of transients is 11,418. Bootstrap sampling errors are shown. The first bin is sensitive to edge effects in the transient detection near the survey boundary; without the $\mathrm{Dec.} > 3^{\circ}$ constraint, the excess in this bin approaches zero. At the highest declinations, the number of detected transients decreases relative to the control sample.}
    \label{fig:declination_excess}
\end{figure*}

Figure~\ref{fig:declination_excess} shows the relative excess of high-probability transients as a function of declination, compared to the control sample. The analysis is restricted to transients satisfying $p > 0.7$, $100^{\circ} < \mathrm{R.A.} < 250^{\circ}$, and $\mathrm{Dec.} > 3^{\circ}$, in order to minimise contamination from plate defects, survey-edge effects, and spatial non-uniformities in the sky coverage.

A significant excess is observed at low declinations ($\sim5$--$20^{\circ}$), where the transient counts exceed the expectations from the control sample by factors of approximately $30$--$75\%$. At intermediate declinations, the excess decreases and becomes negative, while the highest declinations exhibit a relative deficit compared to the control sample. The overall distribution is therefore clearly non-uniform as a function of declination.

The strongest excess occurs at low declinations, corresponding approximately to regions near the celestial equator, where modern optical glints from geosynchronous and high-altitude orbital objects are also preferentially observed \citep{Nir2020}. The result is consistent with a population of objects concentrated near the equatorial plane, where objects in geosynchronous orbits spend most of their time. A second peak is observed near $\sim$65 $^{\circ}$, but the meaning of this finding remains unclear.

\section{Modelling}\label{sec:Results}

\subsection{Orbital Shell Modelling via Earth Shadow Geometry}

In Section \ref{sec:globaldeficit}, we used the global transient shadow deficit to estimate the characteristic orbital altitude of the objects. In this section, we apply a second, independent method of estimating transient altitudes. A simplified analytical framework is used to relate the angular extent of the Earth's shadow to orbital altitude. The angular radius of the shadow region is approximated as a function of Earth radius $R_{\oplus}$ and orbital height $h$, such that higher altitudes correspond to smaller angular shadow regions. This scaling is used to interpret the dependence of the observed deficit on angular distance.

We construct a geometric model of the Earth's shadow, including both umbra and penumbra regions, as a function of angular distance from the anti-solar point. The shadow cones are computed for a range of orbital altitudes, allowing us to predict the expected deficit of observable transients as a function of angle from the anti-solar direction.

A Monte Carlo framework is used to simulate populations of objects distributed within spherical orbital shells at varying altitudes. For each assumed altitude, the expected angular deficit profile is computed and compared to the observed distribution. The altitude corresponding to the observed transition from deficit to no-deficit provides an estimate of the characteristic orbital distance of the transient population. 

We can now compare this model with the empirical results by using the transients and the EarthShadow code to reconstruct the corresponding empirical curve of the Earth-shadow deficit as a function of angular distance. To investigate the orbital altitude associated with the observed Earth-shadow deficit, we performed a $\chi^2$ comparison between the empirical shadow deficit profile and the modelled shadow profiles computed for different shell altitudes. The comparison was restricted to the initial transition, or ``drop'', region of the deficit profile, where most of the altitude-dependent information is contained, rather than the full angular range from $0^\circ$ to $90^\circ$. The analysis was performed using both the complete transient catalogue and the quality-filtered sample and was repeated for three fitting intervals: $0^\circ$--$18^\circ$, $0^\circ$--$20^\circ$, and $0^\circ$--$25^\circ$.

\begin{table*}
\centering
\caption{\textbf{Chi-square comparison between the empirical deficit profiles and the simulated Earth-shadow profiles for different shell altitudes.} Results are shown for the complete and quality-controlled transient samples using three fitting intervals. Here, $\chi^2$ is the chi-square statistic and $\chi^2_{\nu}=\chi^2/\nu$ is the reduced chi-square, where $\nu$ is the number of degrees of freedom. The minimum reduced chi-square value within each fitting interval is shown in bold. Lower X2 values indicate better model fit.}
\label{tab:chi2_altitude_comparison}
\setlength{\tabcolsep}{4pt}
\begin{tabular}{lccccccc}
\hline
& & \multicolumn{2}{c}{$0^\circ$--$18^\circ$}
& \multicolumn{2}{c}{$0^\circ$--$20^\circ$}
& \multicolumn{2}{c}{$0^\circ$--$25^\circ$} \\
\cline{3-4}\cline{5-6}\cline{7-8}
Altitude & Geocentric radius
& $\chi^2$ & $\chi^2_{\nu}$
& $\chi^2$ & $\chi^2_{\nu}$
& $\chi^2$ & $\chi^2_{\nu}$ \\
(km) & (km) & & & & & & \\
\hline
\multicolumn{8}{c}{\textit{Complete transient sample}} \\
\hline
5,000        & 11,371 & 174.51 & 9.70  & 283.48 & 14.17 & 782.17 & 31.29 \\
10,000       & 16,371 & 109.19 & 6.07  & 181.09 & 9.05  & 520.65 & 20.83 \\
15,000       & 21,371 & 69.68  & 3.87  & 115.75 & 5.79  & 277.18 & 11.09 \\
\textbf{20,000}
             & \textbf{26,371}
             & \textbf{63.36} & \textbf{3.52}
             & \textbf{82.75} & \textbf{4.14}
             & \textbf{135.34} & \textbf{5.41} \\
25,000       & 31,371 & 108.73 & 6.04  & 119.97 & 6.00  & 150.16 & 6.01 \\
30,000       & 36,371 & 190.46 & 10.58 & 203.31 & 10.17 & 232.71 & 9.31 \\
GSO (35,786) & 42,164 & 271.81 & 15.10 & 284.66 & 14.23 & 314.07 & 12.56 \\
40,000       & 46,371 & 330.79 & 18.38 & 343.64 & 17.18 & 373.05 & 14.92 \\
50,000       & 56,371 & 435.42 & 24.19 & 448.27 & 22.41 & 477.67 & 19.11 \\
60,000       & 66,371 & 492.94 & 27.39 & 505.79 & 25.29 & 535.20 & 21.41 \\
80,000       & 86,371 & 517.35 & 28.74 & 530.20 & 26.51 & 559.61 & 22.38 \\
\hline
\multicolumn{8}{c}{\textit{Quality-filtered transient sample ($p>0.7$)}} \\
\hline
5,000        & 11,371 & 47.19 & 4.29 & 124.75 & 9.60 & 495.21 & 27.51 \\
10,000       & 16,371 & 30.13 & 2.74 & 91.84  & 7.06 & 391.13 & 21.73 \\
15,000       & 21,371 & 18.15 & 1.65 & 67.31  & 5.18 & 274.99 & 15.28 \\
\textbf{20,000}       & \textbf{26,371}
             & \textbf{13.72} & \textbf{1.25}
             & 46.29 & 3.56
             & 179.90 & 9.99 \\
\textbf{25,000}       & \textbf{31,371}
             & 21.80 & 1.98
             & \textbf{43.94} & \textbf{3.38}
             & \textbf{155.92} & \textbf{8.66} \\
30,000       & 36,371 & 37.87 & 3.44 & 57.65 & 4.43 & 168.97 & 9.39 \\
GSO (35,786) & 42,164 & 48.16 & 4.38 & 67.94 & 5.23 & 179.27 & 9.96 \\
40,000       & 46,371 & 50.07 & 4.55 & 69.85 & 5.37 & 181.17 & 10.07 \\
50,000       & 56,371 & 51.41 & 4.67 & 71.19 & 5.48 & 182.52 & 10.14 \\
60,000       & 66,371 & 56.02 & 5.09 & 75.80 & 5.83 & 187.12 & 10.40 \\
80,000       & 86,371 & 70.97 & 6.45 & 90.75 & 6.98 & 202.07 & 11.23 \\
\hline
\end{tabular}
\end{table*}

Overall, the preferred altitude is remarkably stable across the three fitting intervals. While the best-fitting altitude for the quality-filtered sample shifts slightly from $20\,000$~km for the $0^\circ$--$18^\circ$ interval to $25\,000$~km for the $0^\circ$--$20^\circ$ and $0^\circ$--$25^\circ$ intervals, the complete sample consistently yields a minimum $\chi^2$ at $20\,000$~km above the Earth's surface. This indicates that the shadow altitude inferred from the complete transient catalogue is largely insensitive to the precise definition of the drop region and remains below the GEO/GSO altitude.



We note that the current Monte Carlo framework assumes a simplified random distribution without orbital dynamics or phase-dependent brightness, and is therefore not expected to reproduce all higher-order structure in the observed profile, e.g. the excess near $\sim$47.5$^\circ$.

Taken together, the different estimators favour a characteristic orbital altitude in the range $20\,000$--$35\,000$~km above the Earth's surface. This range reflects the variation between samples, fitting intervals, and altitude estimators rather than a formal statistical confidence interval.


\subsection{Object sizes}

To estimate the object size, we calculate the surface area required to reproduce the observed magnitudes of the transients at altitudes suggested by the previous analyses. In other words, we invert the problem and back-calculate the necessary reflecting area.

To do this, we take into account whether the reflection is diffuse or specular, as well as the exposure time and the assumed altitude of the object. We consider altitudes ranging from 20,000 km to 35,786 km above the Earth’s surface, the latter corresponding to geosynchronous orbit, and assume that the observed signal arises from reflected sunlight. We consider two idealised cases: diffuse (Lambertian) and specular reflection.

We further assume short flash durations of 0.1--1 s, as supported by previous studies \citep{VillarroelARXIV,Busko2026}. These flashes are diluted over the exposure time of the photographic plates (approximately 50 minutes for POSS-I red plates), which is explicitly taken into account in the calculations. The reflectivity is varied between $r = 1$ (perfect reflector) and $r \sim 0.1$ (more realistic, weakly reflective surface).

Purely diffuse reflecting objects (Lambertian surfaces) would not produce short-duration flashes under steady illumination, but instead appear as smeared streaks due to motion across the field during the exposure. Only in the unlikely case of a rapid appearance or disappearance ($\lesssim$1 s) could such objects mimic a flash-like signal. The model therefore assumes a single flash, with no variation in orientation, no inclusion of atmospheric or instrumental effects, and treats all objects as planar reflectors. Under these assumptions, we estimate the effective reflecting area required to reproduce the observed transient magnitudes, assuming perfect reflectivity.

\begin{figure*}
    \centering

    \textbf{(a)}\\[-1mm]
    \includegraphics[width=0.80\textwidth]{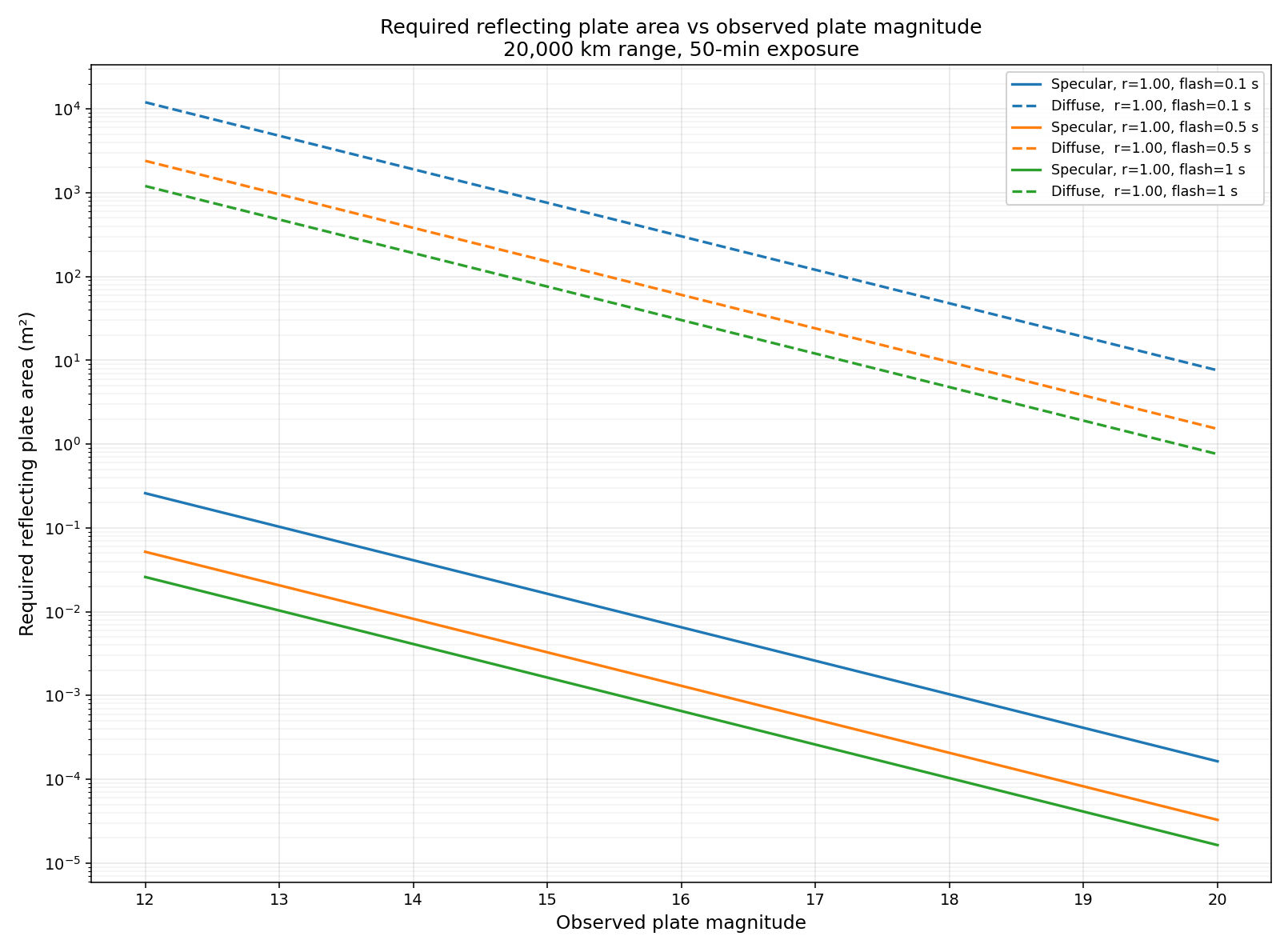}

    \vspace{3mm}

    \textbf{(b)}\\[-1mm]
    \includegraphics[width=0.80\textwidth]{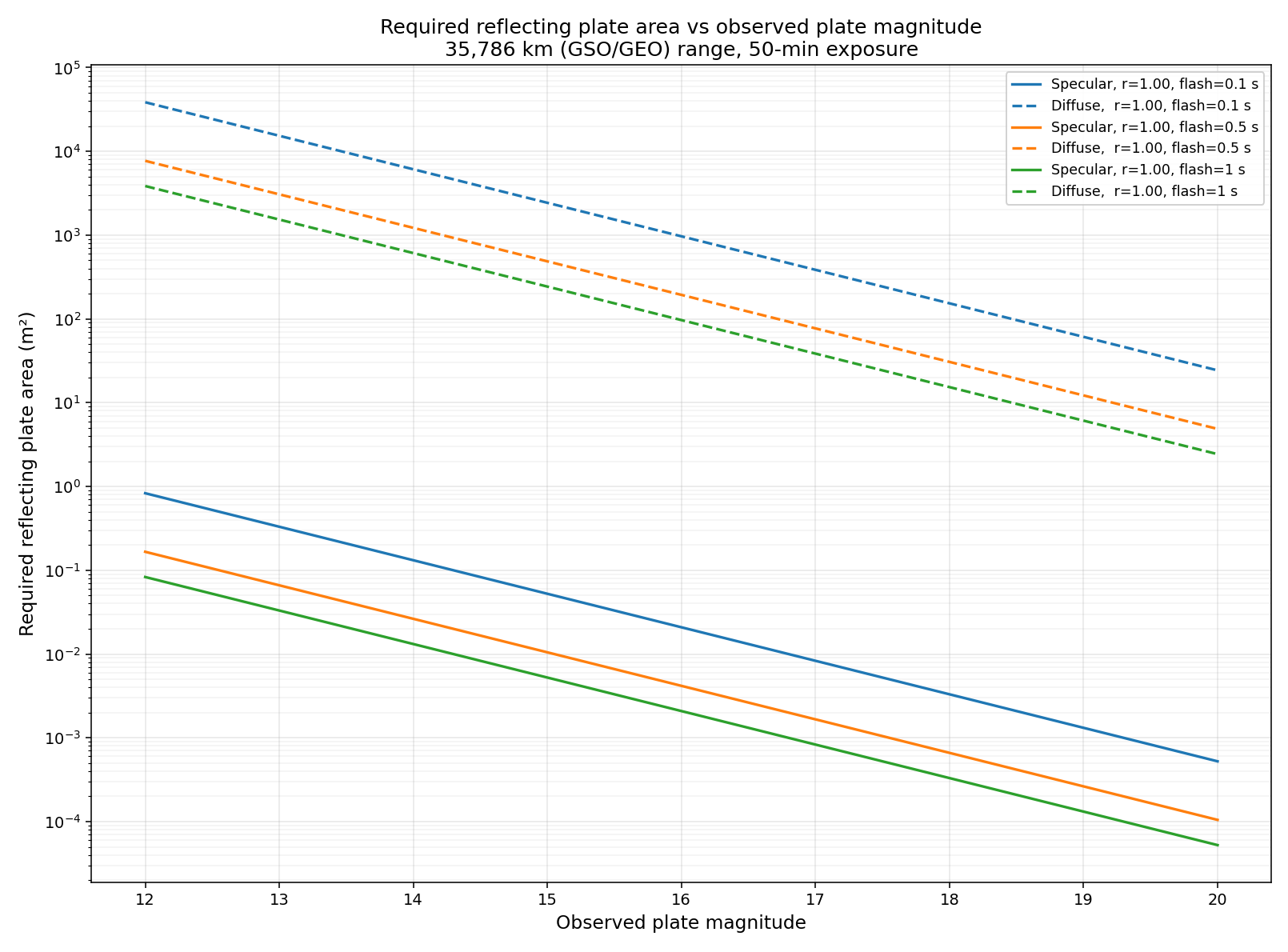}

    \caption{\label{sizes}
    \textbf{Required reflecting-facet area as a function of observed
    plate magnitude.} The inferred reflecting surface area required to
    reproduce the observed transient magnitudes is shown for assumed
    altitudes of (a) $20,000$ km and (b) $35,786$ km (GSO) above the
    Earth's surface. Solid lines represent specular reflection and dashed
    lines diffuse reflection for flash durations of $0.1$, $0.5$, and
    $1$ s, assuming perfect reflectivity ($r=1$) and a 50-min POSS-I
    exposure. Diffuse cases are included for comparison but are not
    expected to produce short-duration flashes under steady illumination.}
\end{figure*}

For this exercise, we compare with representative VASCO examples, such as the five transient cases presented in \cite{VillarroelPASP}, as well as Solano’s triple transient \citep{Solano2024}, to characterise the typical population. While these examples reflect the core sample, we note that many fainter transients exist, and that brighter transients with diffraction spikes are likely underrepresented due to the spike-removal procedures described in \cite{Solano2022}. The results are presented in Figure \ref{sizes}. We include both specular and diffuse cases for completeness; however, diffuse reflection is not expected to produce flash-like transients under steady illumination.

We perform these calculations at both bounds of the inferred altitude range, adopting altitudes of 20,000 km and 35,786 km above the Earth’s surface.

The Candidate 5 example from \cite{VillarroelPASP}, for instance, includes a bright transient with magnitude of $R \sim 12.7$ mag. Assuming a perfect reflector and a flash duration between 0.1--1 s, this corresponds to a planar facet of roughly 0.014 m$^{2}$ up to just over 0.436 m$^{2}$. However, if the surface has low specular reflectivity ($r = 0.1$), for example due to long-term exposure to micrometeorite impacts or cosmic radiation, the required area increases by approximately an order of magnitude and becomes $\sim$0.136 m$^{2}$ to 4.357 m$^{2}$, corresponding to an equivalent linear size of approximately 0.37–2.1 m.

Given that we also detect transients close to the limiting magnitude, and that such objects may plausibly have been exposed to the space environment for hundreds of years or longer, it is more realistic to assume reduced reflectivity. This in turn suggests somewhat larger physical sizes for the underlying objects.

\begin{figure*}
    \centering
    \includegraphics[width=0.80\textwidth]
    {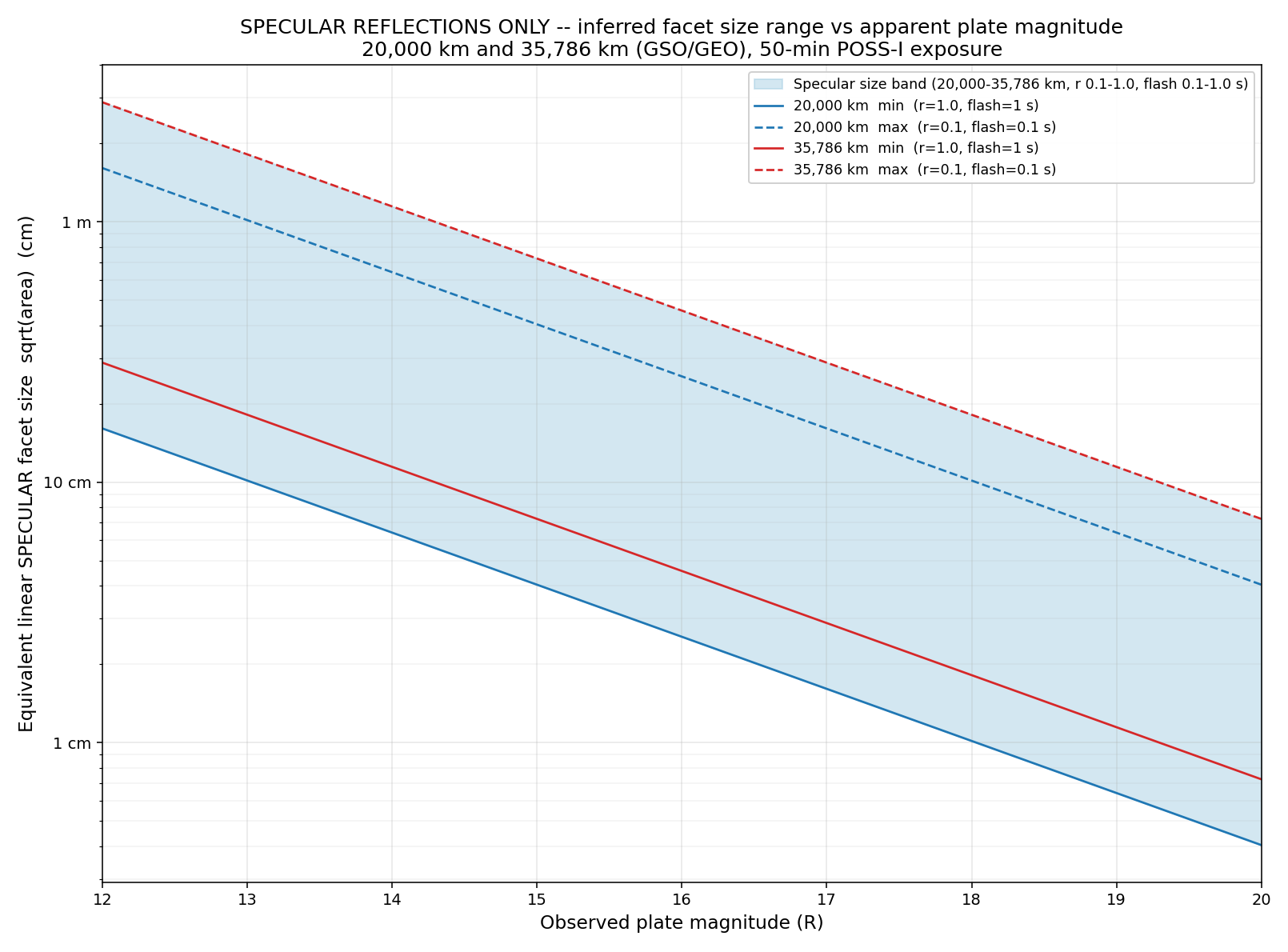}

    \caption{\label{specular_size_band}
    \textbf{Equivalent linear size of specularly reflecting facets as a
    function of observed plate magnitude.} The shaded region shows the
    full range inferred for altitudes of $20,000$--$35,786$ km above the
    Earth's surface, specular reflectivities of $r=0.1$--$1$, and flash
    durations of $0.1$--$1$ s, assuming a 50-min POSS-I exposure. The
    lower boundary corresponds to a perfect reflector ($r=1$) at
    $20,000$ km with a 1-s flash, while the upper boundary corresponds
    to a reflector with $r=0.1$ at GSO ($35,786$ km) with a 0.1-s flash.
    Diffuse reflection is excluded. The full model range extends from
    centimetre scales up to approximately $3$ m.}
\end{figure*}


\subsection{Monte Carlo Simulation of Rotational Rates and Flash Durations}

To constrain the rotational properties of the objects, we do a Monte Carlo simulation that spans all variations on:

\begin{itemize}
   \item object geometry (e.g. cube, icosahedron, flying saucer),
   \item axis rotation
    \item precession
    \item nutation angle,
    \item solar phase angle.
\end{itemize}

We explore various parameter combinations that lead to observe on average $\sim$ 1 flash per plate per object, rather than a pearlband of flashes. We compare two different models, one for cube geometry (6 faces) and another for the icosahedron geometry (20 faces), see Table \ref{tab:cube} and \ref{tab:icosahedron} respectively. We also include a third model corresponding to a saucer-like multi-faceted geometry, characterised by a shallow dome and extended lower hull, as summarised in Table \ref{tab:saucer}.

\begin{table}
\centering
\begin{tabular}{cccc}
\hline
\textbf{Spin [R.P.M.]} & \textbf{Prec [R.P.M.]} & \textbf{Nutation} & \textbf{Flash Duration} \\
\hline
0.03 & 0.05 & 80$^\circ$ & 575 ms \\
0.10 & 0.05 & 80$^\circ$ & 410 ms \\
0.16 & 0.05 & 80$^\circ$ & 205 ms \\
0.03 & 0.10 & 80$^\circ$ & 955 ms \\
0.10 & 0.10 & 80$^\circ$ & 385 ms \\
0.16--0.43 & 0.05 & 85$^\circ$ & 85--250 ms \\
0.20--0.37 & 0.05 & 90$^\circ$ & 60--190 ms \\
\hline
\end{tabular}
\caption{\textbf{Cube geometry.} Simulation results for spin, precession, nutation, and flash duration, using different Revolutions-Per-Minute (R.P.M).}
\label{tab:cube}
\end{table}

\begin{table}
\centering
\begin{tabular}{cccc}
\hline
\textbf{Spin [R.P.M.]} & \textbf{Prec [R.P.M.]} & \textbf{Nutation} & \textbf{Flash Duration} \\
\hline
0.01 & 0.05 & 80$^\circ$ & 1775 ms \\
0.03 & 0.05 & 80$^\circ$ & 1140 ms \\
0.09--0.20 & 0.05--0.10 & 80$^\circ$ & 170--505 ms \\
0.09--0.28 & 0.05--0.10 & 85$^\circ$ & 90--455 ms \\
0.11--0.28 & 0.05--0.10 & 90$^\circ$ & 145--325 ms \\
\hline
\end{tabular}
\caption{\textbf{Icosahedron geometry.} Simulation results for lower spin rates and corresponding flash durations.}
\label{tab:icosahedron}
\end{table}

While there are certainly more possible geometries to test, we can see that the transient objects are likely rotating very slowly (0.01 -- 0.43 R.P.M.), otherwise multiple flashes would be observed during the exposure for most objects. The same constraint applies to extended multi-faceted geometries, such as a slowly tumbling, specularly reflecting saucer-like object at
altitudes of $20,000$--$35,786$ km, where rotation rates of order $\sim1$ revolution per hour ($\sim0.017$ R.P.M.) can produce approximately one to two detectable flashes during a passage through a $6^{\circ}$ POSS-I field.

In such a model and assuming altitudes of $20,000$--$35,786$ km, the instantaneous peak brightness of individual flashes can reach $V \sim 1.0$--$6.3$, i.e. bright enough to be visible to the naked eye under dark skies, while the flash durations remain sub-second.
However, photographic plates record flux integrated over the full exposure time ($T_{\rm exp} \sim 3000$ s), such that short flashes are significantly diluted. The corresponding magnitude dilution is given by $2.5\log_{10}(T_{\rm exp}/t_{\rm flash})$, yielding
$\sim8$--$11$ magnitudes for $t_{\rm flash}\sim0.1$--$1.5$ s. As a result, a bright 1-s flash would be recorded at approximately $V \sim 9.7$--$11.0$, while a fainter 0.1-s flash would appear at $V \sim 16.3$--$17.6$, consistent with the observed transient
magnitude range ($R \sim 12$--$18$ mag).

The objects may consist of simple bodies with a small number of flat reflective surfaces, or more extended objects with predominantly low reflectivity (r $\sim$ 0.1).

\begin{figure*}
\includegraphics[width=\textwidth]{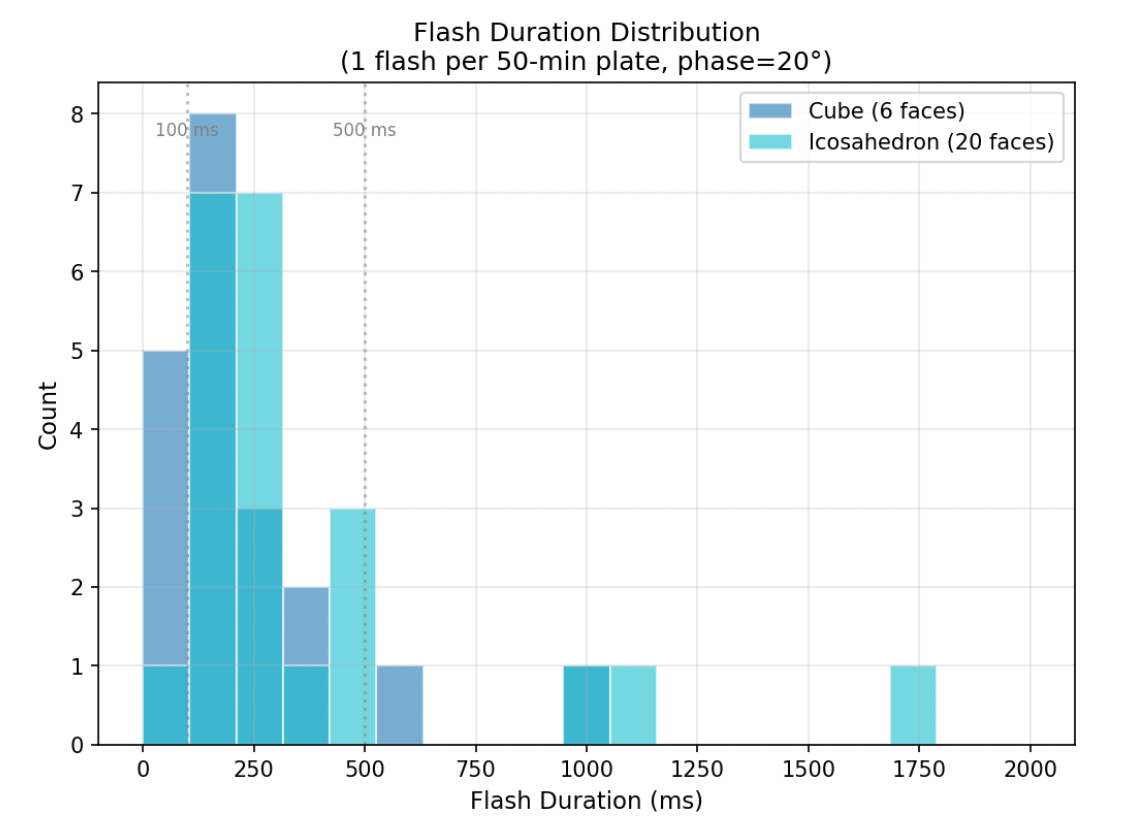}
  \caption{\label{FDD} {\bf Flash Duration Distribution.} Assuming 1 flash per 50-min plate, and a phase = 20°.}
   \end{figure*}

In Figure \ref{FDD} we show the Flash Duration Distribution. Flash durations span $\sim 60$--$1775\,\mathrm{ms}$, with the majority falling within $60$--$600\,\mathrm{ms}$. The combined distribution yields a characteristic flash duration of $\sim 320\,\mathrm{ms}$. However, because of the objects' apparent motion across the sky, flashes longer
than approximately $0.2$--$0.3$ s at $20,000$ km and $\sim0.5$ s at GSO may produce detectable image elongation rather than point-like PSFs.

\begin{figure*}
\includegraphics[width=\textwidth]{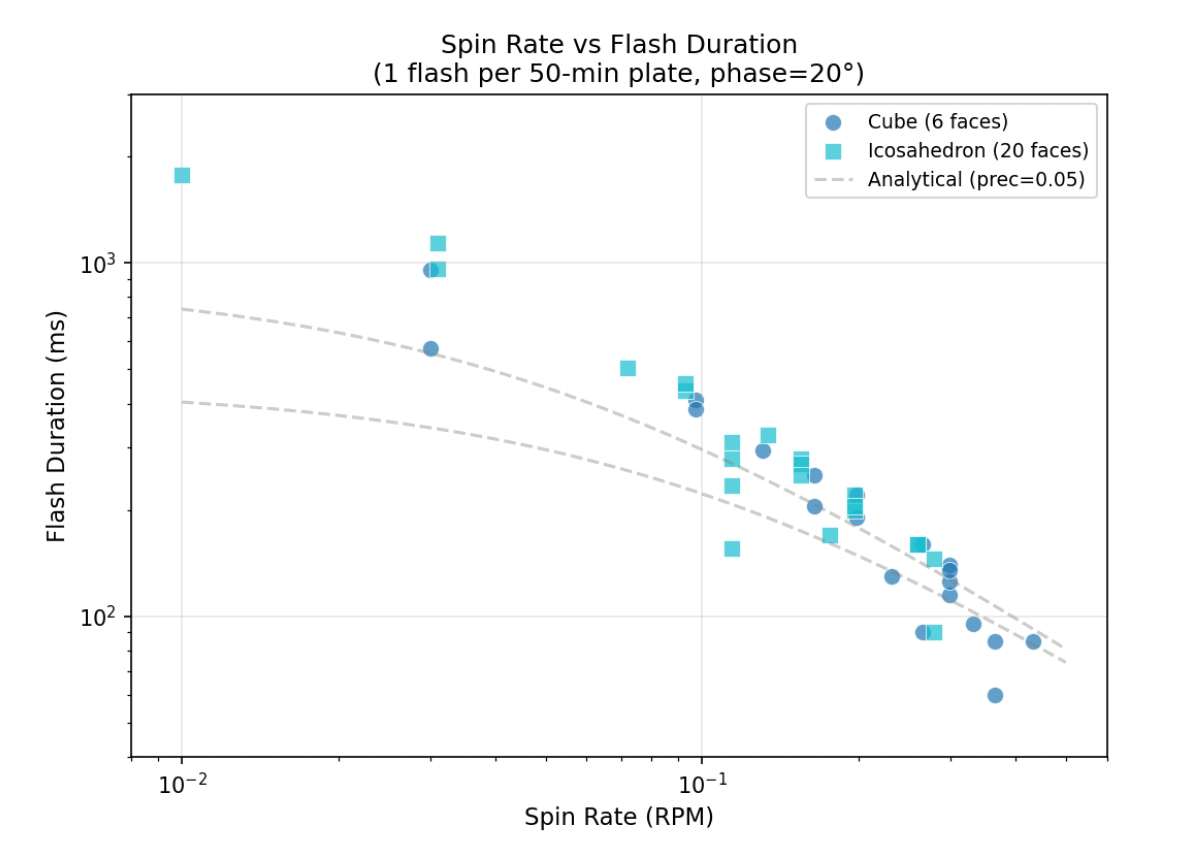}
  \caption{\label{VS} {\bf Spin Rate vs Flash Duration for two Cube and Icosahedron.}}
   \end{figure*}

We also explore the relationship between spin rate and flash duration in Figure~\ref{VS}, comparing cube and icosahedron geometries. As shown in Table~\ref{tab:comparison}, the cube exhibits, on average, higher spin rates and shorter flash durations than the icosahedron.

\begin{table*}
\centering
\begin{tabular}{lcccc}
\hline
 & \textbf{Cube} & \textbf{Icosahedron} & \textbf{Saucer-like} & \textbf{Cube/Icosahedron combined} \\
\hline
Mean s.r. & 0.23 RPM & 0.15 RPM & 0.017 RPM & 0.19 RPM \\
Median s.r. & 0.25 RPM & 0.15 RPM & 0.017 RPM & 0.17 RPM \\
Mean f.d. & 235 ms & 397 ms & 100--1000 ms & 320 ms \\
Median f.d. & 150 ms & 260 ms & $\sim$300 ms & 212 ms \\
\hline
\end{tabular}
\caption{Summary of spin rates (s.r.) and flash durations(f.d. for three different geometries, including a saucer-like multi-faceted model.}
\label{tab:comparison}
\end{table*}

\begin{table}
\centering
\begin{tabular}{lcc}
\hline
\textbf{Region} & \textbf{Flash events} & \textbf{Total light deposit}\\
\hline
Lower hull (underside) & 40 & 3.64 m$^{2}\cdot$s \\
Upper-dome lip & 51 & 0.05 m$^{2}\cdot$s\\
Rim / equator & 0 & 0 \\
Upper dome top & 0 & 0 \\
\hline
\end{tabular}
\caption{\textbf{Distribution of specular flashes across a saucer-like geometry in the simulation.}}
\label{tab:saucer}
\end{table}








\subsection{Velocity constraints}
We do a back-of-the-envelope estimation of the velocity constraints.

For a chain of flashes that fall on a straight line on a photographic plate, the simplest single-object hypothesis is that one moving body flashed periodically as it crossed the field. In that case, the body must have traversed the entire angular extent $\lambda$ within the exposure window $T$, so its tangential speed satisfies
\[
v_{\min} = \frac{\text{range} \times \lambda}{T}.
\]

The slant range is computed for a ground observer at zenith angle $z$ using the law of cosines. Setting $z = 0$ (overhead) minimises the range and therefore the required speed—the most permissive case for the orbital hypothesis.

The orbital column relates to the circular velocity
\[
v_{\mathrm{orb}} = \sqrt{\frac{GM}{r}}
\]
at each reference radius. 

In most cases, the inferred minimum velocities are sub-orbital. For example, the candidates reported in \cite{VillarroelPASP} exhibit angular displacements of approximately 3.6–9.9 arcminutes, corresponding to only a few percent of the plate field of view.
By contrast, the 3–6 degree XE alignments identified in \cite{Doherty2026} span a substantial fraction—up to the full extent—of a photographic plate. Under the assumption that an object traverses the full $6^\circ$ field during a single 50-minute exposure, the corresponding minimum velocity is consistent with the circular orbital velocity at an altitude of approximately 66{,}580 km.
If, however, the same angular displacement is traversed over a shorter interval (e.g., 25 minutes, corresponding to half the exposure time), the implied velocity increases accordingly, yielding consistency with circular orbital motion at an altitude of approximately 41{,}149 km.

These estimates should be regarded as order-of-magnitude constraints. While the lower bounds on velocity are generally sub-orbital, the upper limits remain unconstrained; the objects may plausibly be travelling at orbital velocities or significantly higher.

\subsection{Signs of maneuvering}

A possible interpretation of the observed transient morphology is that the flashes originate from a single object undergoing attitude changes along its trajectory. The data show that multiple point-like flashes often occur along approximately collinear paths, with a tendency for enhanced flash activity at the beginning and end of the track and, in some cases, additional flashes near the midpoint. In a non-negligible fraction of cases, closely spaced ``double'' flashes are observed at the start or end of the trajectory. Notably, these double flashes are separated perpendicular to the apparent direction of motion, rather than along the track itself.

A natural explanation for this behaviour is that the probability of producing a specular reflection depends is sensitive to the rate of change of the object's orientation. For a slowly rotating body, specular alignments between the Sun, the reflecting surface, and the observer are relatively rare. However, during periods of rapid attitude change --- such as those required for reorientation or manoeuvring --- the object's orientation sweeps more quickly through angular phase space, increasing the likelihood of satisfying the specular reflection condition. In this picture, the observed flashes preferentially occur during intervals of enhanced rotational motion.

If the object undergoes directed motion, for example transitioning between trajectory segments, attitude changes are expected both at the beginning and at the end of such segments. This provides a natural explanation for the numerous transient duplets observed in our sample \citep{VillarroelPASP,Doherty2026b}. Occasional intermediate flashes may arise from the underlying slow rotation of the object or from minor attitude adjustments during transit. However, the presence of closely spaced double flashes with a transverse (perpendicular) separation is not readily explained by simple temporal modulation of a single reflecting facet. Instead, it suggests either the presence of multiple reflecting surfaces contributing simultaneously, or a spatially extended or structured reflecting geometry.

The observed angular separation of $\sim$5 arcsec between components in certain double flashes corresponds to a projected physical separation of order $\sim$1\,km at geosynchronous distances. This scale is incompatible with a single compact object or with multiple reflecting facets on a rigid body -- unless it is actually manoeuvring.

While not constituting proof of a manoeuvring object, this interpretation provides a physically consistent framework linking the observed flash distribution to changes in rotational state and orientation, and is difficult to reconcile with purely diffuse reflection, simple rigid-body rotation, or static sources.

\section{Discussion}\label{Constraints}
\subsection{Properties}
Our tentative interpretation places the transient objects at altitudes of $\sim$20,000 -- 35,786\,km above the Earth's surface, consistent with the outer Van Allen belt. The objects likely span a range of sizes depending on the effective
reflecting area of their facets, with estimates ranging from $\sim10^{-4}$ to $\sim8\ \mathrm{m}^{2}$ under different assumptions on reflectivity, flash duration, and observed magnitude. The upper end of this range appears atypical and is associated with bright, short-duration events assuming low reflectivity ($r \sim 0.1$). More representative values, assuming a characteristic flash duration
of $\sim0.3$ s and magnitudes around $R \sim 16$, suggest facet areas
of approximately $0.002$ -- $0.007\ \mathrm{m}^{2}$ for $r=1$
and $0.02$ -- $0.07\ \mathrm{m}^{2}$ for $r=0.1$ across the
adopted altitude range. These areas correspond to equivalent linear
facet sizes of approximately $4.7$--$8.3$ cm for $r=1$ and $15$--$26$ cm for $r=0.1$.

The objects are plausibly composed of a small number of such reflecting
facets (of order a few to tens), and exhibit slow rotation rates with a mean of $\sim0.19$--$0.23$ R.P.M. These rotation rates are consistent with approximately one detectable flash during a passage through the POSS-I field for the geometries explored (cube and icosahedron).

A representative estimate for the majority of transients corresponds
to projected reflecting areas of approximately
0.002$--$0.07 ${m}^{2}$, corresponding to equivalent
linear facet sizes of approximately 5 $-$ 26 cm. For
non-planar objects composed of a limited number of facets, the overall
object dimensions may be larger and depend on the geometry and number
of reflecting surfaces.

\subsection{Possible explanations}
We examine possible explanations for only the Earth-shadow deficit and the observed alignments, initially without taking into account the association between transients and nuclear weapons testing or the transient-geomagnetic storm index anticorrelations. An additional observational constraint is provided by the spatial distribution of the high-probability transients, see \cite{Doherty2026}. While natural Solar-System populations such as asteroids, comets, and zodiacal debris are expected to concentrate preferentially near the ecliptic plane, the Palomar transients instead exhibit a depletion near the ecliptic. We find a pronounced excess at low declinations, corresponding approximately to the celestial equator, see Fig. \ref{fig:declination_excess}. This behaviour is consistent with a population concentrated near the Earth’s equatorial plane, similar to modern high-altitude orbital objects and geosynchronous populations.

We note that the inferred magnitude ranges, sizes, geometries, and flash durations are broadly consistent with transients observed in modern astronomical surveys, which are generally interpreted as specular reflections from human-made satellites and space debris \citep{Nir2020}.

\subsubsection{Natural bodies?}

Conventional interpretations include natural bodies such as reflective ice or debris fragments, as well as small asteroids with locally planar metallic or icy surfaces. Pure ice fragments, however, are unlikely to persist over extended timescales at these altitudes, as sublimation would rapidly reduce their size unless continuously replenished. But short-lived populations of icy cometary debris may be plausible, capable of producing reflections over limited timescales before sublimating away, including as ice on planar surfaces on asteroids, during melting and refreeze cycles.

However, it remains unclear why such material would produce predominantly specular rather than diffuse reflection. Objects dominated by diffuse scattering (e.g. ``dirty'' icy or rocky material) would be expected to generate continuous, low-level illumination over the exposure time, resulting in elongated or non-stellar image profiles. This is inconsistent with the observed short-duration, point-like flashes, and would additionally require substantially larger reflecting areas to reproduce the observed magnitudes. The inferred slow rotation rates ($\sim$0.2 RPM) are not in themselves inconsistent with small objects in space, as rotational states can remain stable over long timescales in the absence of strong torques. However, if the transient population consists of short-lived debris cometary debris, one might instead expect more irregular or rapidly evolving rotational states. The apparent combination of slow, stable rotation and short-lived reflectivity therefore places non-trivial constraints on the physical nature of the objects.

Another hypothesis is that the transients originate from small, highly reflective planar surfaces on asteroids, such as metallic asteroidal fragments. Enhanced reflectivity could arise from recent collisional events, which may expose relatively smooth, planar surfaces capable of producing specular reflections. While surface modification due to the charged particle environment in the outer Van Allen belt cannot be excluded, such processes are more likely to induce surface roughening rather than efficient optical polishing. However, it has been suggested that ion or electron irradiation could in some cases act analogously to beam polishing, preferentially removing softer materials and smoothing surface irregularities at nano- to micro-scales, potentially enhancing specular reflectivity. This remains speculative and depends sensitively on the material composition and irradiation conditions. Such asteroids would also have rough surfaces, which is responsible for diffuse reflection. In such case, we expect a streak to accompany the glint, unless the object is sufficiently small.

We estimate the maximum projected area such diffuse objects can have while still escaping detection (in the form of a streak) in POSS-I. For objects moving across the field during the exposure, the reflected light is smeared over many pixels, reducing the surface brightness and detectability. Using a separate streak model, we find that if the projected area is
smaller than $\sim120$--$170$ square metres (corresponding to
diameters of $\sim12$--$15$ metres), no clearly detectable
streak is expected for an object with low reflectivity ($r=0.1$). This implies that diffuse objects below this threshold may remain undetected in POSS-I due to motion-induced smearing, while still hosting smaller specular substructures capable of producing the observed short-duration flashes. Here, detectability refers specifically to whether small parts of the streak are bright enough to stand out above the local photographic plate background, rather than whether the full trail could be recovered through matched-filter methods or visual inspection of faint streaks. Under more optimistic trail-detection assumptions, even substantially smaller diffuse objects, potentially as small as $\sim$0.8-1.2 m in diameter, could in principle become detectable as faint streaks, thereby significantly weakening natural-body interpretations.

However, asteroids are concentrated near the ecliptic. These objects are avoiding the ecliptic and moreover cluster at the Earth's equator. The observed pattern is therefore inconsistent with what is expected for an asteroid population. Combined with the constraints derived above, the existence of such a naturally occurring population of slow-rotating metallic fragments appears implausible.

Finally, if these transients are the result of the Earth passing through clouds of natural objects, we could expect a temporal correlation between transients and established meteor showers. We searched for correlations between transient detections and annual meteor showers from the IAU Meteor Data Centre using a permutation test that accounts for seasonal observing biases. No statistically robust association was found, and the apparent effect depended on the adopted meteor-shower catalogue.

If the transients are associated with natural bodies, these objects belong to a yet undiscovered population of natural objects.



\subsubsection{Plasmoids?}
More speculative possibilities include novel electromagnetic phenomena in the outer Van Allen belt, potentially modulated by solar and geomagnetic activity. This could explain, for example, the anti-correlation with geomagnetic storm index observed by \cite{Cann2026a,Cann2026b}. A candidate explanation is provided in the same papers by a dusty plasma model operating in the outer magnetosphere. In this scenario, micron- to millimetre-sized charged particles at geosynchronous-like altitudes interact with the ambient plasma environment, with their charging state, dynamics, and radiative properties modulated by geomagnetic activity. During geomagnetic storms, variations in particle flux, energy distributions, and local plasma conditions (e.g. Debye length and sheath structure) may alter the stability, alignment, or emissive behaviour of such particles, leading to a suppression or modification of the observed transient rates. In this picture, the transients are not passive reflectors but manifestations of a magnetospherically coupled plasma process, potentially involving transient charging, electrostatic alignment, or short-lived emission phenomena, see explanations in \cite{Cann2026a,Cann2026b}. It is also conceivable that such ensembles of charged particles could, under favourable conditions, act collectively as a transient specular reflector. In this scenario, a momentary alignment of particle orientations and positions could produce a brief, mirror-like flash toward the observer, even if the individual constituents are small and irregular. However, the millimeter-scale particle sizes of such plasmas do not align well with the inferred object sizes derived in this work. Unless the present size estimates are substantially revised by future modelling, this discrepancy essentially rules out dusty-plasma interpretations of the transient population.

\subsubsection{Non-human intelligence (NHI)?}

While one may consider a population of small natural asteroids (with linear sizes below 12-15 m depending on the altitude) containing at least one highly planar, reflective metallic surface, such a model can at best account for the Earth-shadow deficit. It does not naturally explain the temporal correlations with reported UFO events and nuclear tests \citep{Bruehl2025}, or the anticorrelation with the geomagnetic storm activity index \citep{Cann2026a,Bruehl2025}. If the transient population instead represents a heterogeneous mixture of events with different natural origins, some partial solutions may be found within the natural toy models.

However, if the transients share a common origin, any viable explanation must simultaneously account for all of these additional observational features, while also being consistent with the clustering of objects near the celestial equator (and the deficit near the ecliptic). Among the toy models we consider for this paper, an NHI scenario is the only one that, in principle, simultaneously accommodates all five observational features: (1) PSFs with, on average, slightly smaller FWHM than real stars, (2) observed groupings, (3) a deficit of transients in the Earth's shadow, (4) a correlation in time with nuclear testing (and UFOs), and (5) an anticorrelation with the geomagnetic storm index Kp. That being said, nature can always surprise us with new physical phenomena beyond our current imagination.

The possibility that the transients arise from specular reflections of artificial objects associated with non-human intelligence has been explored in several works by the VASCO team (see e.g. \citealt{Villarroel2022a,Bruehl2025,VillarroelPASP,Bruehl2026}). Such a framework naturally explains not only the absence of reflections within the Earth's shadow, but also the temporal association with nuclear weapons testing (e.g. in the context of possible planetary-scale monitoring), as well as the anticorrelation with geomagnetic storm activity. In addition, the frequent occurrence of closely spaced doublets within the transient sample are potentially indicative of controlled attitude changes or manoeuvring. Given the existing observations, including that the transient–nuclear correlation is stronger for sunlit objects \citep{Doherty2026}, the data favour a common origin over a mixed-population scenario.

While the small inferred sizes may at first appear perplexing, it is worth noting that Breakthrough Starshot has proposed deploying thousands of gram-scale, wafer-sized nanocraft, each coupled to a metre-scale lightsail, on a potential flyby mission to the Alpha
Centauri system. This provides a concrete modern example of how large populations of extremely small artificial probes could be deployed for interstellar exploration.

\section{Future work}\label{sec:Future}

Taking all of the above into account, the next step is to improve the results and define where to focus and how to improve the analysis. We recommend the following directions for further work:

\begin{enumerate}
    \item Reflectance / reflectivity improvements. Adopt a more physically realistic surface-scattering model by replacing the perfect-mirror assumption with a bidirectional reflectance distribution function (BRDF). This approach would enable constraints on surface properties (e.g.\ polished metal, paint, ice) through the relationship between brightness, beam broadening, and surface roughness, and would refine estimates of object size and reflectivity.
    \item Refining and validating the modelling pipeline. A key next step is to calibrate the simulations against well-characterised modern reflective objects by applying the same forward model to sources with known properties, such as Iridium antenna glints, Starlink reflections, and GPS satellites at Medium Earth Orbits (MEO) and GSO altitudes. Reproducing their observed magnitudes, durations, and event rates would anchor the model in real data and help identify any systematic biases in the photometry or geometric assumptions. In parallel, the modelling can be further constrained by applying a streak-morphology test to the observed transient FWHMs. For flashes lasting longer than $\sim$1~s, the simulations predict measurable motion (of order 4--20 arcsec, which would produce elongated streaks rather than point-like sources. Comparing these predictions with the observed point-spread functions of the transients provides an independent means of testing and constraining slow-tumbling regimes.
    \item Perform Monte Carlo simulations using empirically motivated priors on rotation and tumbling parameters, rather than uniform assumptions. These priors can be informed by observed tumble rates of deorbited debris, where available. By combining the per-object specular-flash model with these distributions, we can directly predict the plate-level distributions of transient counts and brightness for comparison with the observations. This enables a direct comparison between simulated and observed transient populations, and helps constrain the number of objects required and the dynamical behaviours needed to reproduce the measurements.
    \item To isolate the contribution of anthropogenic objects, one can perform a differential comparison between historical and modern transient catalogues with matched selection functions. By applying consistent cuts (e.g.\ magnitude limits, morphology, and sky coverage) and normalising to a common rate per unit area and time, the transient distributions as a function of angular distance from the antisolar point, $f(\theta)$, can be directly compared. The difference, $\Delta f(\theta) = f_{\mathrm{modern}}(\theta) - f_{\mathrm{historical}}(\theta)$, then traces the time-evolving component of the population, expected to be dominated by human-made objects. Any residual component sharing the same functional form as the historical distribution would indicate a persistent subpopulation. Forward modelling of the historical sample under modern survey conditions can further account for differences in exposure time and sensitivity.
    \item Using modern surveys similar to that presented in \cite{Nir2020}, we can specifically search for duplets of transients at the beginning and end of an alignment, oriented perpendicular (or near perpendicular) to the overall direction of motion of the aligned transients. These searches are best conducted near the equatorial plane, where most Palomar transients are found, as well as the majority of transients detected in modern surveys. These objects would be of particular interest as potential examples of maneuvering sources. With simultaneous observations from two telescopes, it would also be possible to estimate the altitude of an individual object through parallax and assess changes in its velocity and direction of motion. A possible approach is to use a modern transient catalogue similar to that presented in \cite{VillarroelMNRAS}, but constructed outside the Earth's shadow, and to search it for analogous features indicative of maneuvering objects. The exposure time of the Zwicky Transient Facility will, however, unintentionally include many groups of asteroids in the catalogue, and particular care will therefore be required to exclude them.

\end{enumerate}

Together, these steps ensure that the model is both physically realistic and consistent with the observational signatures. Additional approaches, such as spatial analyses of aligned multiple transients, may provide further constraints.

\section{Conclusions}\label{sec:Conclusions}

In this paper, we present the first attempt to constrain the properties of transients detected on photographic plates from the first Palomar Observatory Sky Survey (POSS-I), for which the most plausible interpretation, is short flashes or glints from objects near the Earth. We use geometric shadow modelling, Monte Carlo simulations, and photometric constraints to infer their physical properties. We examine the observed groupings and alignments of multiple transients, which appear consistent with sudden changes in direction. The measured deficit of transients in Earth’s shadow as a function of angular distance from the antisolar point is consistent with a population located at altitudes of $\sim 20,000 - 35,000$ km (extending into geosynchronous altitudes).

We model a range of geometries, from simple standard shapes to more complex, highly multifaceted configurations. While more complete and detailed simulations are still required, the first results suggest that we are dealing with very slowly rotating reflective structures with non-smooth geometries, such as cubes, icosahedron or flying saucer-like objects, rather than smooth, spherical surfaces (e.g. spheres). Also, more complex geometries are currently considered unlikely given the observed transient count per plate, unless additional effects — for example more representative reflectance modelling (BRDF), or extremely slow tumble — reduce the effective per-plate flash rate to a level consistent with observation.

Across the adopted altitude range of $20,000$--$35,786$ km, our modelling suggests reflecting facets with equivalent linear sizes ranging from centimetre scales up to approximately $3$ m, depending on the observed magnitude, flash duration, and specular reflectivity.

We note a dependence of inferred size on altitude, with the inferred characteristic sizes decreasing toward lower altitudes. Our simulations suggest that the population is best described as a non-homogeneous mixture of small metallic flat objects, ranging from centimetre scales up to approximately 3 meters, and objects exhibiting relatively simple geometries (i.e., a low number of reflecting facets). A limited population of larger fragments may also be present, contributing to the observed diversity in transient morphologies and magnitudes.

Interestingly, these inferred properties of the objects from our modelling -- namely, slowly rotating flat objects with reflecting facets ranging from centimetre scales up to approximately $3$ m --  appear remarkably similar to the size and behaviour of the sources observed by Redstone-Mercury and Apollo mission astronauts. \footnote{See e.g. https://www.war.gov/UFO/\#NASA-UAP-D2-Apollo-17-Transcript-1972.} We speculate that some of the objects observed by Apollo astronauts may have been similar metallic surfaces. In this scenario, the same population could already be present in the 1950s data, but was subsequently interpreted as material released from the rocket or spacecraft.

An important limitation of this study is that it relies on the sample selection of \cite{Solano2022}, in which bright objects were excluded. As a result, larger objects in geosynchronous orbits were filtered out, biasing our sample toward smaller objects and potentially affecting our conclusions. In addition, streaks and elongated objects were excluded because our analysis focused exclusively on point sources. Consequently, our study may miss both larger objects and a population of fast-moving objects in lower orbits. To have been detected by the Palomar survey and to have survived the rigorous filtering applied by \cite{Solano2022}, specularly reflecting flat surfaces had to be sufficiently small to remain fainter than the bright-source threshold; sources with $R \leq 12.4$ mag were excluded. In fact, this paper provides a practical demonstration of how unknown objects could be routinely rejected by the quality-control procedures used in most modern astronomical transient surveys. This broader methodological problem was previously highlighted by \cite{VillarroelKrisciunas2024}.

Finally, we suggest a simple strategy for detecting similar transients in contemporary data by searching for duplet transients oriented perpendicular to the direction of the alignment, indicative of maneuvering objects. Although this is challenging given the large amount of human-made debris and satellites in orbit, there remain opportunities to distinguish these candidate sources from known anthropogenic signatures. Brief transients with similar flash durations, complex shapes, and comparable apparent magnitudes are routinely observed in modern astronomical surveys, where they are generally attributed to specular reflections from human-made satellites and space debris at geosynchronous altitudes, see e.g. \cite{Nir2020}. It is therefore reasonable to ask whether a significant fraction of modern ``satellite glints'' may actually originate from a population of objects already present in the pre-Sputnik era. It is possible that at least some present-day Uncorrelated Targets and unexplained transient optical tracks detected in modern surveys may be associated with this unknown population of objects.

\section{Acknowledgments}

The authors express their deep gratitude to N. Colosimo for the extensive technical input, codes, calculations, and conceptual insights that were instrumental to this study. B.V. further thanks C. Franck for support. B.V. is funded by the Swedish Research Council (Vetenskapsr\aa{}det, grant no. 2017-06372) and is also supported by an anonymous donor, to whom she is deeply grateful. A.S. thanks the Athanatos Foundation for support.

\bibliographystyle{plainnat}
\bibliography{aa_cit}


\label{lastpage}
\end{document}